\documentclass[a4paper,10pt]{scrartcl}
\usepackage[utf8x]{inputenc}
\usepackage[numbers]{natbib}
\usepackage{graphicx}
\usepackage{listings}
\usepackage{hyperref}
\hypersetup{%
	colorlinks,
	linkcolor = black,
	pdfpagemode = UseNone
}

\graphicspath{{./images/}}

\title{Enhanced TKIP Michael Attacks}

\author{Martin Beck, TU-Dresden, Germany\\
\textless martin.beck1@tu-dresden.de\textgreater}
\date{February 25, 2010}

\begin{document}

\maketitle

\begin{abstract}

This paper presents new attacks against TKIP within IEEE 802.11 based networks. Using the known Beck-Tews attack, we define schemas to continuously generate new keystreams, which allow more and longer arbitrary packets to be injected into the network. We further describe an attack against the Michael message integrity code, that allows an attacker to concatenate a known with an unknown valid TKIP packet such that the unknown MIC at the end is still valid for the new entire packet. Based on this, a schema to decrypt all traffic that flows towards the client is described.

\end{abstract}

\section{Introduction}
\label{sec:intro}

IEEE 802.11\citep{IEEE07} is a standard family for wireless networks. Such networks can be found in home, office, and enterprise environments and
are quite popular today. If sensitive information is transmitted over a wireless network, privacy and integrity is a concern
and must be taken care of.

Following the first generation of IEEE 802.11 WEP encryption, which has major design flaws and was broken several times, beginning with
a complete break by Fluhrer, Mantin, and Shamir in 2001 \citep{FMS01}, WPA with its ``TKIP'' protocol was introduced as an intermediate
successor of WEP on the way to the final IEEE 802.11i amendment. Since its 2002 introduction, TKIP was believed to be secure untill
in late 2008 a first attack was published. This attack recovered the Michael key towards the wireless client and thus allowed an attacker to
generate a valid MIC code for an arbitrary packet and send that as a genuine frame to the client.

The structure of this paper is as follows: In Section~\ref{sec:beck_tews}, we give an introduction to the
technical details of the Beck-Tews attack. In Section~\ref{sec:new_keystreams}, we lay out two schemas on how to generate new keystreams
including additional requirements and obstacles to overcome. In section~\ref{sec:reset_attack}, Michael is analyzed and a simple way
of generating collisions is presented. Based on these collisions which set the internal Michael state to an arbitrary value, a
key reset attack is developed, that in the end allows for packet concatenation. In Section~\ref{sec:conclusion}, our findings are
summarized and mitigation techniques suggested.

\section{Beck-Tews Attack}
\label{sec:beck_tews}

In november 2008 an attack was published by Beck and Tews \citep{BT08}, which took the WEP chopchop attack \citep{Kor04}, switched the
target from the accesspoint to the client, used ``michael error report frames'' for validating a correct guess, instead of looking out
for replayed packets and implementing a 60 seconds timeout between each byte to defeat the countermeasure initiation. It works the same
as the original chopchop attack, going backwards over the whole packet, guessing the last byte until the right guess validates the ICV
and triggers the ``michael error report frame'', as the MIC is invalid with a very high probability. The attack utilizes an arp packet,
as most of its content is known or could easily be guessed, so that the overall number of unknown bytes is as small as possible.

It takes at least twelve rounds to get the ICV and MIC bytes, which are at the end of each packet and cannot be guessed in advance.
After completion, the keystream for the choped packet is know and thus the plaintext including the MIC and the ICV\@. Using the reverse
Michael algorithm presented in \citep{BT08}, the MIC key for that direction is calculated and can now be used to generate valid MICs.

As the IEEE 802.11e amendment is used to reuse already transmitted IVs on remaining QoS channels, the recovered keystream can encrypt
up to seven new packets, one on each remaining channel. With a recovered keystream length of 40 bytes, the forged packet and its use
is very limited. As the packet must include the encrypted 4 byte ICV and 8 byte MIC, only 28 bytes can be used for the plaintext packet.

\section{Generating new keystreams}
\label{sec:new_keystreams}

The goal of sending longer packets can be achieved through fragmentation at the IEEE 802.11 layer. Support for up to 16 fragments is
included since the first version of the wireless protocol. The design of TKIP prevents the use of a single IV-keystream pair for more
than one fragment. So in order to send 16 fragments, 16 different keystreams are needed. As the discussed attack takes over 15 minutes
to complete for an arp frame, it would not be applicable to generate new keystreams that way.

Most packets in IEEE 802.11 wireless networks are either ARP or IPv4 packets and carry a LLC header in front. This LLC header and the
beginning of the following ARP/IP header can be guessed and is known most of the time, allowing a passive collection of short
keystreams. The LLC header is 8 bytes in length and completely known for either ARP and IPv4 packets. For following ARP packets,
the content is known up to the last bytes of the source IP address and can be used together with the generated MIC and ICV to rapidly
get the full ARP keystreams \citep{OM09}.

The IP packet header starts with a version and length tag, set to $0x45$, or $0x46$ in case there is an IP option field
used. The ``Differentiated Services Field'', defined by $0x00$ or $0xc0$ most of the time is the second byte that can be guessed. Given
the total length of the transmitted packet, the following two bytes can be calculated, as they carry this value. So for IPv4 packets,
a total of 12 keystream bytes are known and can be collected for nearly every packet that arrives.

While the 4 byte ICV value applies at the MPDU level, every fragment needs to carry it, whereas the 8 byte MIC is calculated over the
whole MSDU frame body and is therefore only present in the last fragment. As we get to know 12 keystream bytes, 4 bytes down for each ICV
, eight bytes can be used to encrypt an arbitrary packet that spans over 16 fragments. As the last fragment needs to carry the MIC,
an attacker can send packets with a size up to 120 bytes, just by listening to passing IP packets. In case ARP packet keystreams are
used, this yields a maximum packet length of 568 bytes. ICV and MIC bytes are already removed, as the complete keystream length would
have been 640 bytes.

\begin{figure}[htbp]
  \centering
    \includegraphics{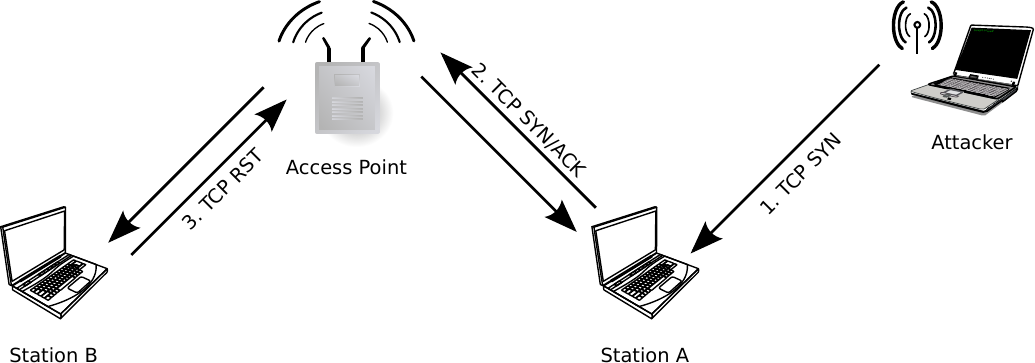}
  \caption{local TCP Scan}
  \label{figure:tcp_scan_local}
\end{figure}

Once the ARP packet is decrypted, the IP range and at least one IP address is known. Using the fragmentation attack described above to
encrypt spoofed TCP-SYN packets towards a clients open TCP Port, the client will send a TCP-SYN/ACK in return to the spoofed IP address.
The spoofed address may be the wireless access point for example, which should answer with an TCP-RST packet and thus generates a new
IV towards the client. This schema is shown in~\ref{figure:tcp_scan_local}.

The forged packet consists of the 8 byte LLC header, the 20 byte IP header and the 20 byte TCP header, which adds up to 48 bytes of
MSDU length. After computing and adding the MIC, 56 bytes of keystream are needed in order to encrypt the packet. As each short
keystream allows us to encrypt 8 bytes of plaintext, 7 fragments are needed to send the TCP-SYN packet. As each sent packet generates
exactly one response and can be sent on the remaining 7 channels, there is no gain in using this method, as we never get more usable
keystreams.

However, many Linux systems send a zero ID within the IP header of TCP-RST packets and many wireless access points are
running linux, so in that case an attacker gains the two extra keystream bytes from listening to TCP-RST packets, followed by the flags
and fragment field. As there are no other fragments, this is always zero and the flags would be either $0x40$ for a ``don't fragment''
flag set, or zero otherwise. The next ``TTL'' field is often set to $0x40$ as a default by linux systems and the protocol on top of IP
will be TCP, as we specified that, which gives a $0x06$ on that byte. The source and destination IPs are also known, as we chose them
initially, so the IP header checksum can be calculated and inserted. Now we got the complete IP header keystream. The TCP header is also
completely known to the attacker, as the ports were chosen in the first place, the sequence number will be the same as the original
incremented by one and the ack number will be zero. Header length and flags are always set to $0x50$ and $0x04$ describing a 20 bytes
TCP header and the RST flag being set. The window- and urgent field will be zero in TCP-RST frames and the Checksum can again be
computed as all necessary fields are known.

So in case the attacker finds the IP of a local linux system, which has the described features, the complete TCP-RST packet can be
guessed, including the MIC and ICV bytes, which generates a new (8 bytes LLC, 20 bytes IP, 20 bytes TCP, 8 bytes MIC and 4 bytes ICV)
60 bytes keystream, which in return can be used alone to encrypt a TCP-SYN packet, that generates up to 7 new 60 bytes keystreams
without any fragmentation needed.

\begin{figure}[htbp]
  \centering
    \includegraphics{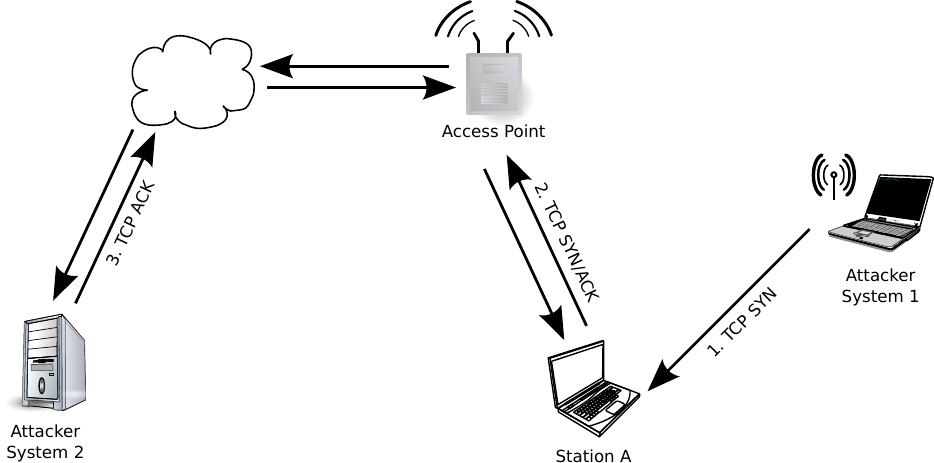}
  \caption{remote TCP Scan}
  \label{figure:tcp_scan_remote}
\end{figure}

So if these requirements are met, an attacker can generate new 60 byte keystreams, at the speed of the wireless network connection.
However if there is no such system available, the source IP can be spoofed to meet the IP address of a remote system, that is controled
by the attacker. This remote scenario is shown in~\ref{figure:tcp_scan_remote} and gives the ability to exactly specify the content and
size of the final TCP handshake packet. This packet will be known to the attack, up to the value of the ``TTL'' field, which can be
guessed and will stay within a close range of possible values. So again the complete keystream is known, as the plaintext was chosen by
the attacker and can even have a padding, which gives additional keystream bytes. This enables the attacker to generate arbitrary length
keystreams at the speed of the inbound WAN connection. This requires the attacked wireless network to have a connection to the internet
and the attacked client to be configured accordingly.

\section{Michael Reset Attack}
\label{sec:reset_attack}

In \citep{BT08} a reverse Michael algorithm was defined, which is used to calculate the MIC key, used to generate a MIC for a given
plaintext. The initialization of Michael sets the two Key words as the internal state from which all following 32 bit words are then
processed. So if the internal state of the Michael algorithm reaches a point, where the two internal words have the exact same value
they had at the beginning, after the Michael Key was set, the remaining plaintext will have the same MIC value, as the whole packet.

\begin{figure}[htbp]
  \centering
    \includegraphics{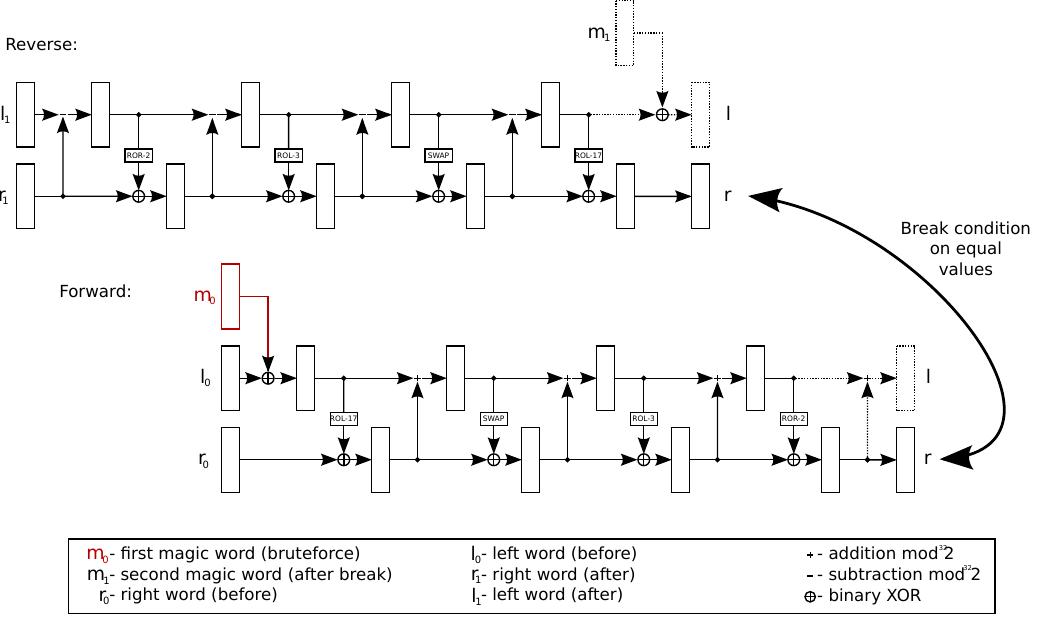}
  \caption{Calculation of the magic words}
  \label{figure:magic_words_calc}
\end{figure}

So if it would be possible to reset the internal state to the initial values, all the bytes in front would have no influence on the
final MIC\@. Given a known plaintext we want to append at the front of an arbitrary and unknown packet together with the known MIC Key
for that direction the internal state after inserted plaintext can be calculated. The internal state after our custom plaintext and
the internal state just before the original packet is know, which is the same as the Michael Key. Figure~\ref{figure:magic_words_calc}
shows how two calculate two magic words, which will fit in between the custom plaintext and the original packet to reset the internal
state of the Michael algorithm.

To calculate these magic words, an attacker takes the final state and calculates backwards using the reverse Michael algorithm
from \citep{BT08} up to the point where the second magic word would be XORed into the state. At this point the value of the right
internal word is known and will be kept as a reference. It is the same value that will be calculated using the first magic word, which
is still unknown and applying the original Michael algorithm on that.

At this point, there is no further computation possible, as we either need the second magic word to continue calculating the reverse 
Michael algorithm, or the first magic word to begin with the ordinary algorithm. For the next step, all possible values for the first
magic word will therefore be bruteforced, giving a maximum total number of $2^{32}$ Michael rounds. Once the right word of the internal state
reaches the reference value we stored after the first step, the bruteforce is stopped and the first magic word is found.

Once this is done, which takes $2^{31}$ Michael rounds on average, all that is left is the calculation of the second magic word. Using
the two dotted operations in figure~\ref{figure:magic_words_calc}, the second magic word is calculated by applying the operations left on
the left word of the internal state. This gives the second magic word. Once the algorithm finished, two 32-bit words, the ``magic words''
are calculated, which reset the internal state of the Michael algorithm.

\begin{figure}[htbp]
  \centering
    \includegraphics{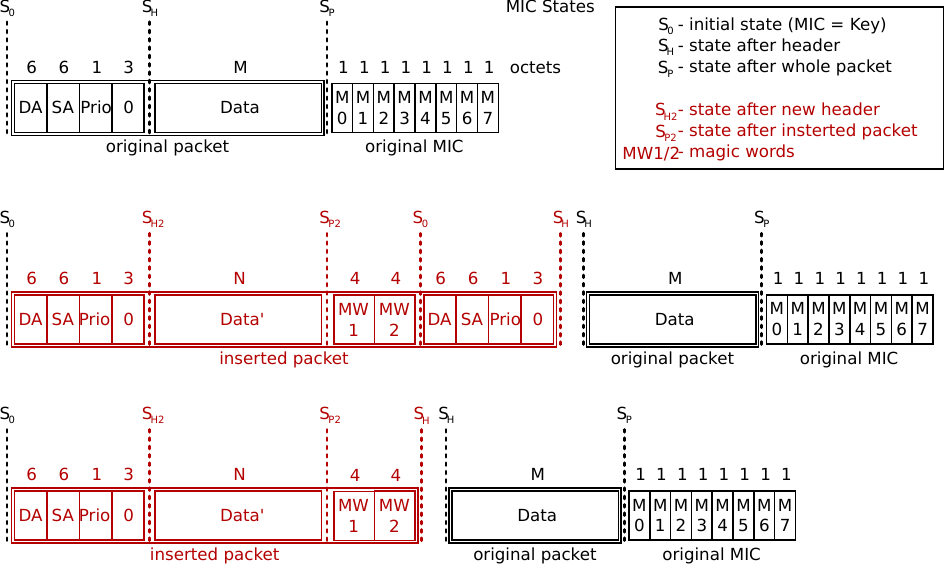}
  \caption{Packet concatenation}
  \label{figure:michael_reset_attack}
\end{figure}

This can be used to concatenate an arbitrary known plaintext packet with an unknown plaintext packet and keeping the unknown MIC at
the end of the original packet valid for the whole new packet. Figure~\ref{figure:michael_reset_attack} denotes the structure of the
concatenation. The original packet will be set as the second fragment and will stay exactly the same, except for the fragment and QoS
channel field. These fields will not be included in the MIC calculation on the second frame and thus will not invalidate the MIC.

The first frame will have an arbitrary plaintext, for which the two magic words are calculated and appended. After these two words
the internal state is set back to the Michael Key. Calculation of the MIC also includes the Source Address, Destination Address and
Priority fields of the IEEE 802.11 header, which need to be included too, as they will only be included from the first fragment. They
need to have the exact same values which were present in the original frame, so that the internal state is changed the same way it was
originally. This adds another 16 bytes after the two magic words and completes the first fragment.

Another way to calculate the two magic words, is to not get to the state, where the Key was set, but just after the 16 header bytes were
processed. This removes the necessarity to include these 4 words after the two magic words and thus saves 16 bytes of keystream.
The downside with that method is, that the magic words for a fixed inserted plaintext, depend on the value of the Source address and
original priority. If they change, the magic words need to be recalculated. The whole algorithm to calculate the magic words takes
about 45 seconds on a 1.7 GHz Pentium-m and about 15 seconds on a dual core T4400 processor. It can easily be split upon multiple
threads, by crawling through different parts of the bruteforced space.

\begin{figure}[htbp]
  \centering
    \includegraphics{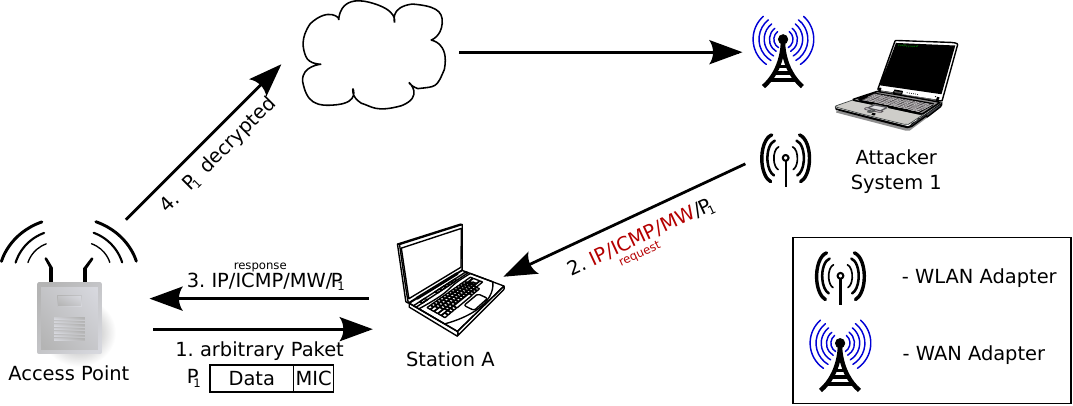}
  \caption{Decryption through ICMP echo request}
  \label{figure:remote_icmp_decryption}
\end{figure}

As a result we define an attack, which takes any captured packet, changes it to match an appropriate QoS channel and to be the second
fragment. The first fragment will be a complete IP/ICMP echo request packet, with a destionation IP address of the attacked client and
a source IP address of the WAN port of the attacker. This WAN connection can be established through a UMTS connection or another IEEE
802.11 wireless network available. So if ICMP echo requests are allowed at the client side, it will generate an ICMP response towards
the WAN address. The response will carry the same padding that was sent to the client through the original ICMP echo request and is
in our case set to the two magic words, maybe the 16 header bytes and the compelte original packet.

Once this reponse arrives at the WAN port, the entire unknown packet can be extracted from the padding and is therefore decrypted.
Additionally the complete keystream of that packet is known and can be used to inject other packets, or insert ICMP packets infront of
other received frames to decrypt them. Using this technique, any packet towards the client can be decrypted within a rather short amount
of time. The ICMP packet itself can span over several fragments, as it was described earlier for TCP-SYN packets.

\section{Optimization}
\label{sec:optimization}

To find a Michael collision for a given input and ouput state pair, the whole domain for either the first or second magic word will be
checked untill a collision is found. It is possible that not exactly one collision for a given pair, but more or even none exists.
The first case, there exists more than one solution, will help speeding up the search, whereas the second case, with the absence of a
solution, will result in a failed attack.

Building upon the first case, the probability for reaching a collision at the beginning of the domain will be increased if there is more
than one solution. To apply that to our algorithm, lets assume there would be several possible and acceptable Michael states for the packet
that is to be inserted in front of the existing packet, there would also be several magic words to get a collision for any of the given
states. To get these different states, the inserted packet is altered and each variation results in a new Michael state.

So now we have a set of packets, where each is good enough to be inserted before the original packet and thus a set of Michael states,
generated out of the set of packets. Also we have the final Michael state, which needs to be reached by processing the magic words
upon any of the input Michael states. It doesn't matter which packet and thus state is chosen, as all are equally good in terms of
functionality.

The collision algorithm needs to be changed in a way, that no longer the first magic byte gets bruteforced and checked against the
right word of the Michael state just between the magic words, as this requires the initial state to be constant, as it is used as the
base for the bruteforce. So if the second magic word is iteratively set to every possible value and upon that the reverse Michael
algorithm is applied again. The resulting right word of the Michael state can be checked against the whole set of initial Michael states
or their right words to be exact.

This allows to check the set of initial states for each iteration of the bruteforce, increasing the probability to find a collision
earlier in the domain. To avoid checking the entire set by applying a comparison against each of the containing states, a filter is
generated to group several states together, that allows for only checking the calculated state against a subset of the initional states
in case the filter matches.

To generate this filter an iterative approach is used, which goes over the last $n$ bits of each right word in the set of initial states
and calculates the number of odd bits for the current bit position. In case there are more odd than even bits, all states and thus packets,
which carry an even bit on this position are removed and the filter is set to a 1 on the current position. This is repeated for all $n$
bits, generating a subset of the previous set of initial Michael states by removing about half the number of remaining states for each
iteration. If there are more even bits, the filter gets set to 0 on that position and all words with an odd bit on the current position
are removed.

\begin{figure}[htbp]
  \centering
    \includegraphics{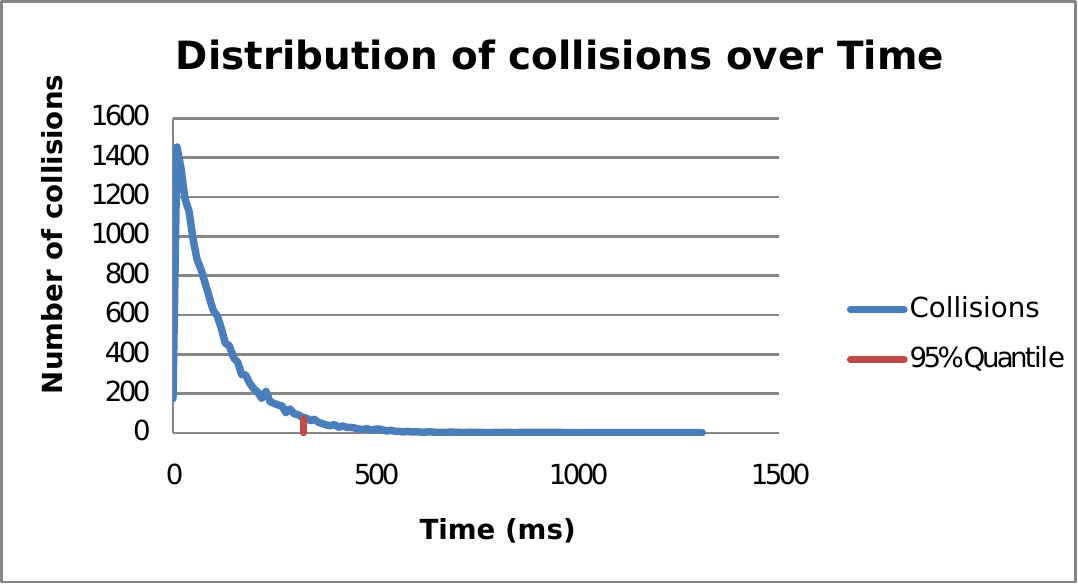}
  \caption{First collisions distributed over Time}
  \label{figure:michael_collision_e4600}
\end{figure}

The number of initial states would be $2^k$, where $k>n$. After all this is done, an $n$ bit filter and a subset of the initial states with at
least $2^{n-k}$ states remains, which will be used while crawling through the domain. So for every iteration during the bruteforce,
the last $n$ bits of the resulting right word are checked against the previously generated filter and in case they match, all states of the
subset are checked against this calculated state.

\begin{figure}[htbp]
  \centering
    \includegraphics{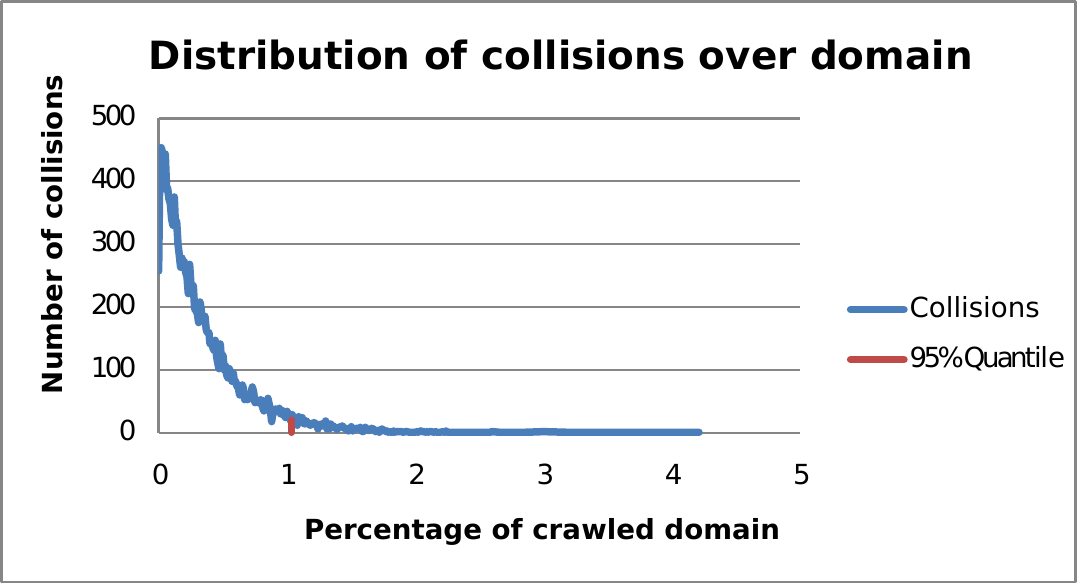}
  \caption{First collisions distributed over the crawled domain}
  \label{figure:michael_collision_e4600_2}
\end{figure}

Now, depending on the number of bits used for the filter and the number of packets used for the set of initial Michael states, the
bruteforce does only one comparison for on average $2^n-1$ iterations, which would be the check of the filter. Once the filter matches,
there will be at least $2^{n-k}+1$, but probably some more, comparisons including the filter and the complete subset. This leads to a
collision finding algorithm, which is a lot faster than the original version and has a much higher probability of finding a collision at
all. Figure~\ref{figure:michael_collision_e4600} shows the distribution of first collisions over time on an intel E4600 processor using $n=8$,
$k=16$ and $2^12$ different Michael keys to find the first collision for the given variables. With a 95\%-quantile at 323 milliseconds and
an average of 77 milliseconds to find the first collision, this is more than two magnitudes faster than the original algorithm.
Figure~\ref{figure:michael_collision_e4600_2} denotes the distribution of first collisions over the domain of the second magic word. This
sets the 95\%-quantile to 1.04\% and the average to 0.24\%. This means that on average after checking 0.24\% of the $2^{32}$ possible values
in the bruteforced domain an collision is found.

In the attack described in this paper, inserting an IP/ICMP header before the original packet, the set of different packets can be filled
by taking the final packet that is to be inserted and setting all possile values for the IP id field. As the id field has a width of 16
bits, $2^{16}$ different packets will be generated. If that is not enough, the ICMP header also carries a 16 bit id field and a 16 bit sequence
number value, which can be chosen arbitrarily and add by itself another $2^{32}$ possible packets.

\section{Conclusion}
\label{sec:conclusion}

The early attacks on TKIP were using a dictionary attack against the initial four way eapol handshake to try to get to know the pre
shared key, which is slow and has a very low success probability, but attacks the master key which is used for everything and
 - if successful - is a complete break. The Beck-Tews attack \citep{BT08} presented in 2008 only targets the access point to client
communication, retrieves the MIC Key for that direction, allowing validation of packets and injection of a very few and short new
packets. Injection and decryption of packets towards the client is now possible without the previous limitation, arbitrary packets
can be sent and decrypted.

The requirements are still the same or even tighter than for the original attack, as we assume either an existing connection to the
internet, a linux system as part of the wireless network, an open TCP port on the attacked wireless client or allowed ICMP echo frames
from remote IP addresses. Sometimes even some of these assumptions are implied together.

The attack can therefore be prevented the same way it is possible for the original attack. Deactivating QoS or setting the rekeying
timout to a low value like 120 seconds. The preferred way is to disable TKIP and switch to the more secure CCMP instead, as most current
devices need to support this protocol suite to be compatible to the IEEE 802.11i amendment, which is now part of the 2007 version of the
standard.

\bibliographystyle{plainnat}
\bibliography{./literatur.bib}

\end{document}